\documentclass[aps,prb,showpacs,reprint]{revtex4-1}

\usepackage{graphicx}

\date{\today}

\begin{document}

\title{Nonclassical properties of electronic states of aperiodic chains in
a homogeneous electric field}

\author{B.J. Spisak}
\email{spisak@novell.ftj.agh.edu.pl}
\affiliation{School of Physics and Astronomy, University of Leeds,
 Leeds LS2 9JT, United Kingdom}
\affiliation{Faculty of Physics and Applied Computer Science, AGH University of
 Science and Technology, al. Mickiewicza 30, 30-059 Krak\'ow, Poland}

\author{M. Wo{\l}oszyn}
\email{woloszyn@agh.edu.pl}
\affiliation{Faculty of Physics and Applied Computer Science, AGH University of
 Science and Technology, al. Mickiewicza 30, 30-059 Krak\'ow, Poland}


\date{\today}

\begin{abstract}
The electronic energy levels of one-dimensional aperiodic systems driven by a
homogeneous electric field are studied by means of a phase space description
based on the Wigner distribution function.
The formulation provides physical insight into the quantum nature of the
electronic states for the aperiodic systems generated by the Fibonacci and
Thue-Morse sequences.
The nonclassical parameter for electronic states is studied as a function of the
magnitude of homogeneous electric field to achieve the main result of this work
which is to prove that the nonclassical properties of the electronic states in
the aperiodic systems determine the transition probability between electronic
states in the region of anticrossings.
The localisation properties of electronic states  and the uncertainty product of
momentum and position variables are also calculated as functions of the electric
field.\footnote{Published as: B.J. Spisak, M. Wo{\l}oszyn
\href{http://dx.doi.org/10.1103/PhysRevB.80.035127}{Phys. Rev. B \textbf{80} (2009) 035127}.
Copyright (2009) by the American Physical Society.
}
\end{abstract}

\maketitle

\section{\label{sec:level1}Introduction}

Analysis of the electronic states of artificial structures (e.g. superlattices,
quantum wires, rings or dots) plays a central role in modern condensed matter
physics because they determine many useful properties indispensable in
industrial applications of atomic- or nano-scale devices.
The experimental realisation of that kind of structures allows to investigate
the influence of quantum effects on mechanical, optical or transport
properties.
The atomic cluster structures such as quantum corrals or chains can be
fabricated using the low temperature scanning tunnelling microscope (STM)
manipulation of individual atoms on the conducting substrate \cite{Eigler1990,
Meyer2000} or mechanically controllable break junction (MCBJ)
\cite{Muller1992,Yanson1998,Smit2001}.
The latter method allows to fabricate only very short mono or mixed atomic
chains that consist of 4 or 5 metal atoms \cite{Ohnishi1998,Bettini2006}; on
the contrary, the tip of STM can be used to build longer and much more complex
structures with different shapes.
Additionally, long atomic chains can be reconstructed on a flat surface by
self-assembly.
Experimental results for self-assembled gold atomic chains on silicon surfaces
confirmed the existence of long and stable chains \cite{Crain2004,Krawiec2006}.
In this case the chains are extended over hundreds of nanometres and can break
due to atomic defects or because of the intentional removal of single atom or
group of atoms from the perfect chain.
It means that experimental techniques open up the possibility to fabricate novel
metal nanostructures created by the intentional atomic rearrangement.
On the other hand, the external fields interacting with electronic states of the
systems can modify their electronic spectrum and new properties of the system
are observed.
Especially the interaction with the electronic states in the system with broken
translational symmetry (disordered or aperiodic systems) seems to  be
interesting  because the quantum interference effects determine the electronic
properties of the systems at low temperatures ~e.g. \cite{qts:lifszyc, Lee1985,
kramer1993}.
One of the simplest but nontrivial examples of the interactions is the effect of
the homogeneous electric field on electronic states in crystals where the
Wannier-Stark quantisation was predicted~\cite{Wannier1960, Avron1982,
Nenciu1991, Grecchi1995}, and  confirmed  experimentally in artificial
semiconductor and optical superlattices ~\cite{Mendez1988, Voisin1988,
Ghulinyan2005}.
Discussion of the effect in one-dimensional periodic systems has been carried
out for many years \cite{Shockley1972, Fukuyama1973, Roy1982, Soukoulis1983,
Prigodin1989, Ryu1993, deBrito1995, Farchioni1997, Gluck1998} and is still the
subject of current experimental as well as theoretical research
\cite{Rosam2001, Gluck2002a, Kast2002, Zhang2006}.

In this paper, we apply the phase space approach based on the non-classical
distribution functions \cite{Tatarskii83, Hillery84, Lee95, Schleich2001} to the problem of
electronic states in isolated and finite one-dimensional aperiodic systems
generated by the Fibonacci and Thue-Morse sequences \cite{Luck1989, Bovier1993,
Hu2000} in the presence of an external homogeneous electric field.
This approach is widely used in the study of quantum transport phenomena
\cite{Morgan1985, Frensley1990, Biegel1996, Jacoboni2004, Wojcik2008}, but less
exploited in the description of electronic states of nanosystems
\cite{Spisak2007}.
The phase space method allows one to carefully investigate
the increasing role of the quantum effects resulting from
interplay between the electric field and aperiodic ordering in the
one-dimensional systems and leading to the greater importance of the
nonclassical properties of electronic states.
We compute the Wigner distribution functions for some electronic states in the
aperiodic systems and analyse their localisation properties in the phase space
as a function of electric field.
The analysis is based on the inverse participation ratio (IPR) in the phase
 space~\cite{Ingold2003, Woloszyn2004}.
This approach allows us to show that the electric field increases the
nonclassical properties of electronic states close to  the transition between
them.

The presented analysis can be simply extended to the description of cold atoms
in the optical lattice \cite{Wilkinson1996, Madison1999, Billy2008} or layered
systems such as aperiodic superlattices consisting of two different materials
that are arranged according to appropriate aperiodic sequence, e.g.
\cite{Perez-Alvarez2001, Gluck2002b, Macia2006} and references therein.

The rest of the paper is organised as follows. In Sec. \ref{sec:level2}, we
present the theoretical model of pseudoatomic chain.
In  Sec. \ref{sec:level3} we apply the formalism of non-classical distribution
functions to the quantitative analysis of electronic states in the aperiodic
systems in the external electric field.
In concluding remarks we summarise presented results.

\section{\label{sec:level2}Theoretical model and methods}

We consider the aperiodic chain of metallic potential wells modelling atoms
which can be described by the one-particle Hamiltonian in the form
\begin{equation}
\mathcal{H}_0=\frac{p^2_x}{2m}+\sum_{i=1}^Nv(x-X_i),
\label{eq:ham}
\end{equation}
where $m$ is the effective mass of electron, and $N$ is the number of wells.
The positions of wells, $X_i$, are distributed according to the binary Fibonacci
and Thue-Morse sequences.
We generate these sequences over set $\{0,1\}$ using the following inflation
rules \cite{Bovier1993}:
\begin{enumerate}
\item[a.] Fibonacci sequence: $0\longrightarrow 01$, and $1\longrightarrow 0$,
\item[b.] Thue-Morse sequence: $0\longrightarrow 01$, and $1\longrightarrow10$.
\end{enumerate}
In our notation zero corresponds to an empty site in the simple crystal lattice,
and one corresponds to a site in the lattice occupied by the potential well
$v(x-X_i)$.

The potential term in the Hamiltonian (\ref{eq:ham}) is represented by the
superposition of the potential wells $v(x-X_i)$.
We assume that each well in the chain is given by the Shaw pseudopotential
modified by screening, namely
\begin{equation}
v(x-X_i)=-v_0
\left\{\begin{array}{ll}
\frac{\exp{(-\mu|x-X_i| )}}{|x-X_i|}, \; |x-X_i|>x_c\\
\frac{\exp{(-\mu x_c )}}{x_c}, \; |x-X_i|\leq x_c.
\end{array}\right.
\label{eq:pspot}
\end{equation}
A quantity $x_c$ is the cut-off parameter and it is chosen to have the value for
which the pseudopotential reproduces the ionisation energy of Cu; $\mu$ is the
screening factor in the Thomas-Fermi approximation.
We assume that each well gives only one state to the conduction band and
 therefore the Fermi level, $E_F$, is defined in the middle of the conduction
 band.
Finally a constant electric field $\mathcal{E}$ is applied along the wire and
the total Hamiltonian of the system under the electrostatic perturbation has a
form
\begin{equation}
\mathcal{H}=\mathcal{H}_0+e\mathcal{E}x,
\label{eq:totham}
\end{equation}
where $-e$ is the electron charge.

\begin{figure}
\includegraphics[width=\columnwidth]{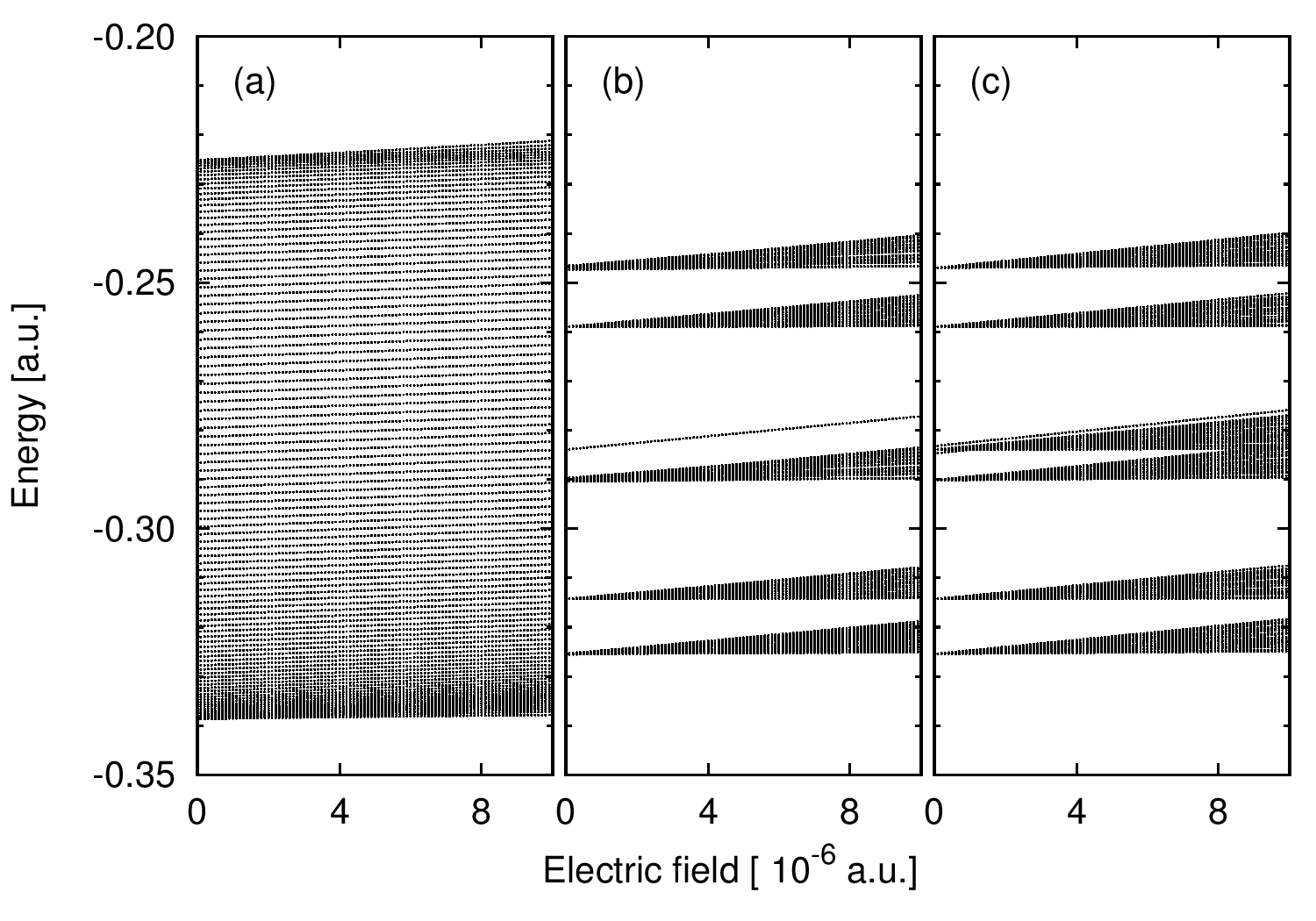}
\caption{Energies of states in the conduction band for N=100 wells as a function
of the electric field for (a) periodic, (b) Fibonacci and (c) Thue-Morse
sequence.}
\label{fig:01}
\end{figure}

\begin{figure}
\includegraphics[width=\columnwidth]{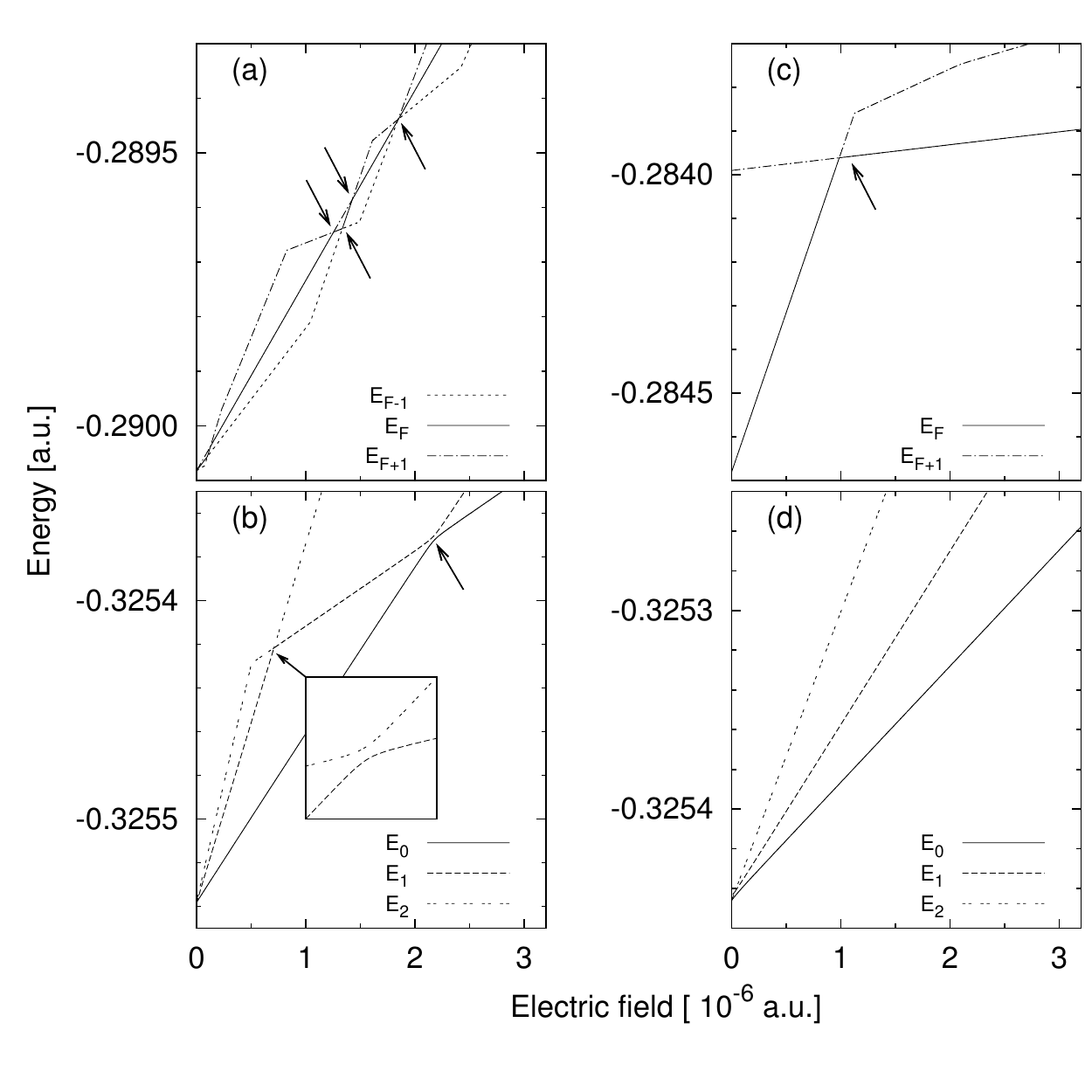}
\caption{
Energies calculated for (a), (b) the Fibonacci sequence and (c), (d)
Thue-Morse sequence.
(a) and (c) show states at the Fermi level ($E_{F}$) and directly below and
 above the Fermi level ($E_{F-1}$ and $E_{F+1}$, respectively).
(b) and (d) show states at the bottom of the conduction band.
Arrows point to the anticrossings.
}
\label{fig:02}
\end{figure}

The energy spectrum of the finite system which is described by the Hamiltonian
(\ref{eq:ham}) forms the energy bands.
The structure of these energy bands strongly depends on the arrangement of
 potential wells.
When a periodic system is subjected to the electric field, the eigenstates of
the Hamiltonian (\ref{eq:totham}) form so-called homogeneous Wannier-Stark
ladders \cite{Wannier1960, Wannier1962}.
This situation is presented in Fig. \ref{fig:01}. a.
In the aperiodic structures generated by the Fibonacci and Thue-Morse order, the
eigenstates of the Hamiltonian (\ref{eq:totham}) form more complex energy
spectra of the conduction band as it is shown in Figs. \ref{fig:01}. b and
\ref{fig:01}. c.
In these cases, the electronic states tend to group into subbands.
The energy  widths of these subbands are much smaller than the width of the
conduction band for the periodic system and strongly depend on the values  of
electric field and number of wells in the systems.
As a result, the aperiodic systems form the inhomogeneous Wannier-Stark ladders.
From these results we conclude that the effect of inhomogeneous Wannier-Stark
ladders originates from the interplay between aperiodic order and the electric
field on the electronic states.
A closer inspection of the inhomogeneous Wannier-Stark ladders reveals the
occurrence of changes in energy levels inclinations in the electric field.
These changes result from the repulsion between neighbouring electronic states,
and the anticrossings~\cite{Davydov1965} are observed between them for some
values of the electric field as it is shown in Fig. \ref{fig:02}.

These non-trivial properties of the inhomogeneous Wannier-Stark ladders suggest
that the correlated position and momentum behaviour of electronic states play
an important role in the region of anticrossings.
Therefore the nature of these electronic states is analysed by the phase-space
methods.
Here, we restrict ourselves to the detailed analysis of the states in the
vicinity of the band bottom, but we also present briefly some results for
the states in the middle  of the conduction band.

Within the phase space approach, the electronic states may be represented  by
the Wigner distribution function.
For an electron in the pure state the Wigner distribution function has the form
\cite{Tatarskii83, Hillery84, Lee95}
\begin{equation}
f_n(x,k)=\int{\mathrm{d} x^{\prime}}\langle
x-\frac{1}{2}x^{\prime}|\psi_n\rangle\langle\psi_n|x+\frac{1}{2}x^{\prime}
\rangle \mathrm{e}^{ikx^{\prime}}.
\label{def:WDF}
\end{equation}
It should be noted that the Wigner distribution function is a bilinear
combination of the electron wavefunction and therefore contains interference
information.
By means of the Wigner distribution function we can evaluate the expectation
value of any Hermitian quantum-mechanical operator $\mathcal{A}$ in the state
$n$ using the formula \cite{Tatarskii83, Lee95}
\begin{equation}
\langle A\rangle_n=\int{\mathrm{d}x \mathrm{d}k} \, A(x,k) \, f_n(x,k),
\label{eq:A_n}
\end{equation}
where $A(x,k)$ is the Wigner representation of quantum-mechanical operator
$\mathcal{A}$ given by
\begin{equation}
A(x,k)=\int{\mathrm{d}x^{\prime}}\langle
x-\frac{1}{2}x^{\prime}|\mathcal{A}|x+\frac{1}{2}x^{\prime}\rangle
\mathrm{e}^{ikx^{\prime}}.
\end{equation}
Because the Wigner distribution function can take negative values in some
subregions of the phase space it cannot be interpreted as the classical
distribution function in the phase space.
The negative part of the Wigner distribution function is responsible for quantum
correlations between spatially separated pieces of the electronic
state~\cite{Zurek1991}.
It stems from the fact that the information from the off-diagonal terms
in~(\ref{def:WDF}) represented by $x^{\prime}\neq 0$ variable is transferred to
the Wigner distribution function via the momentum $k$.
These properties characteristic for the Wigner distribution function may be
utilised as an indicator of nonclassicality of electronic states
\cite{Benedict1999, Kenfack2004}.

\section{\label{sec:level3} Results and discussion}

In the present considerations we assume that the aperiodic systems generated by
the Fibonacci and Thue-Morse sequences are limited by the infinite wells.
The energy spectra were calculated for weak electric fields using the parameters
 of the conduction electron in copper.
In our calculations we consider chains of different lengths (e.g. 30, 50 or 100
 potential wells).
Here we present the results for 100 wells containing all characteristic
 properties of the obtained results.
All values are given in atomic units $\hbar=|e|=m_e=1$.

\begin{figure}
\includegraphics[width=\columnwidth]{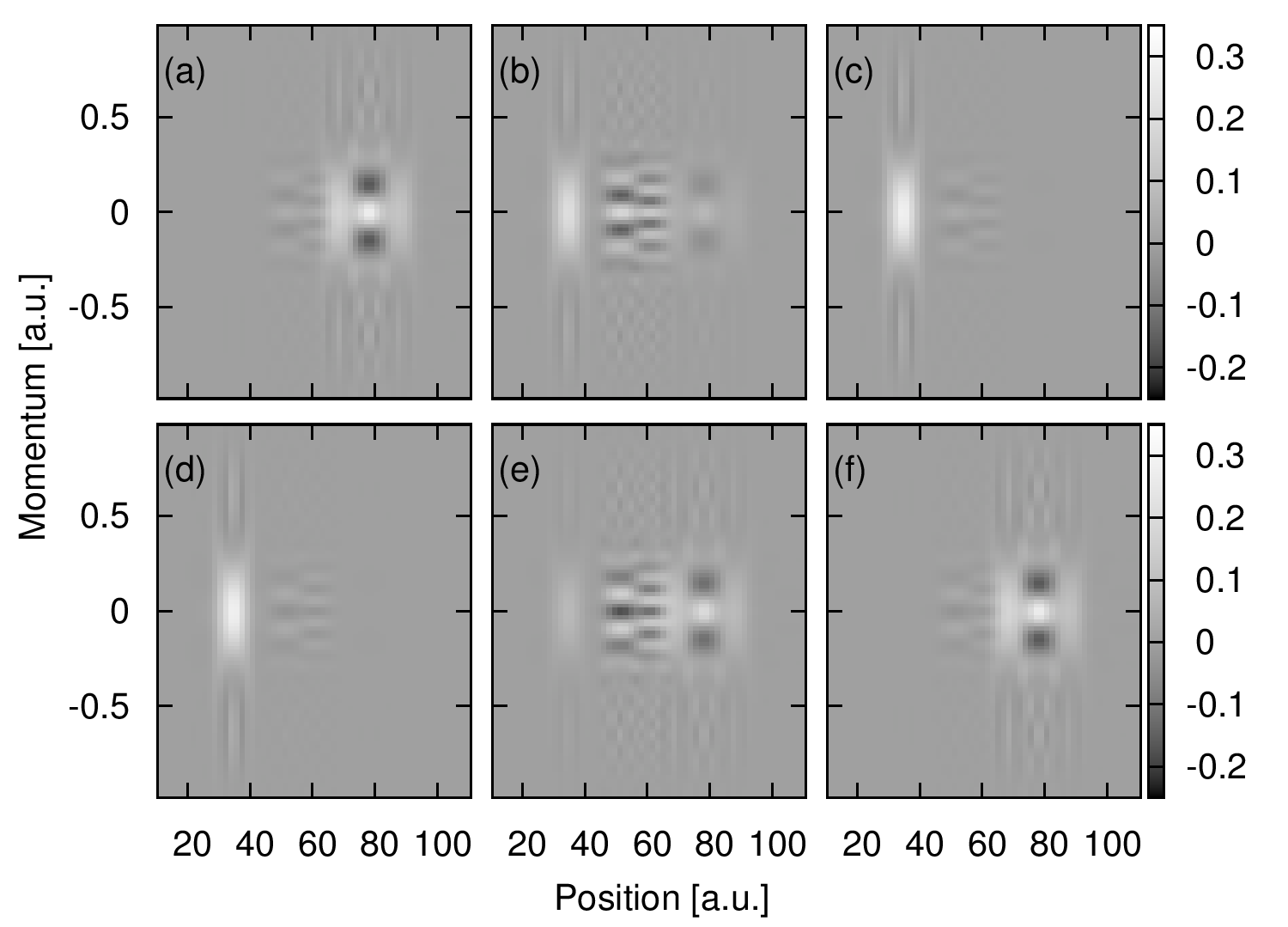} 
\caption{The Wigner function for the system based on the Fibonacci sequence.
(a), (b), (c) -- n=0; (d), (e), (f) -- n=1. Electric field, from left to right,
is equal 1.9, 2.2, 2.5  $\times$ 10$^{-6}$ a.u.}
\label{fig:04}
\end{figure}

In the first step we determine the Wigner distribution functions for some of
pure states and investigate their changes due to the electric field.
Fig. \ref{fig:04} shows the Wigner distribution functions for the two lowest
electronic states in the Fibonacci chain in the vicinity of the first
anticrossing that is pointed to by the arrow in Fig. \ref{fig:02}. b.
Initially, both electronic states which are represented by the Wigner
distribution functions $f_0(x,k)$ and $f_1(x,k)$ occupy different regions of the
phase space.
Increasing the electric field shifts the states in opposite directions so that
the distance between them decreases.
Finally, in the region of anticrossing, the Wigner distribution functions occupy
the same region of the phase space, and it leads to overlapping of these
functions  as shown in Figs \ref{fig:04}. b and \ref{fig:04}. e.
In this case we may express the electronic states by the linear superposition of
individual states: $|\psi\rangle=a_0|\psi_0\rangle+a_1|\psi_1\rangle$, where
$|a_0|^2+|a_1|^2=1$.
Using the definition given by Eq. (\ref{def:WDF}) we obtain expression for the
Wigner distribution function of the electronic state in the form
\begin{eqnarray}
f(x,k) & = & |a_0|^2 f_0(x,k) + |a_1|^2 f_1(x,k) \\
& + & 2\Re \bigg\{ a_0 a_1^{\ast} \int{\mathrm{d}x^{\prime}} \langle
x-\frac{x^{\prime}}{2}|\psi_1\rangle\langle\psi_0|x+\frac{x^{\prime}}{2}
\rangle \mathrm{e}^{ikx^{\prime}}\bigg\}.\nonumber
\label{interf}
\end{eqnarray}
The first two terms  correspond to individual electronic states, and the last
term represents the mixture part which results from the bilinearity of the
Wigner distribution function.
This type of term is often termed the quantum interference \cite{Dowling1991,
Agarwal1995}.

\begin{figure}
\includegraphics[width=\columnwidth]{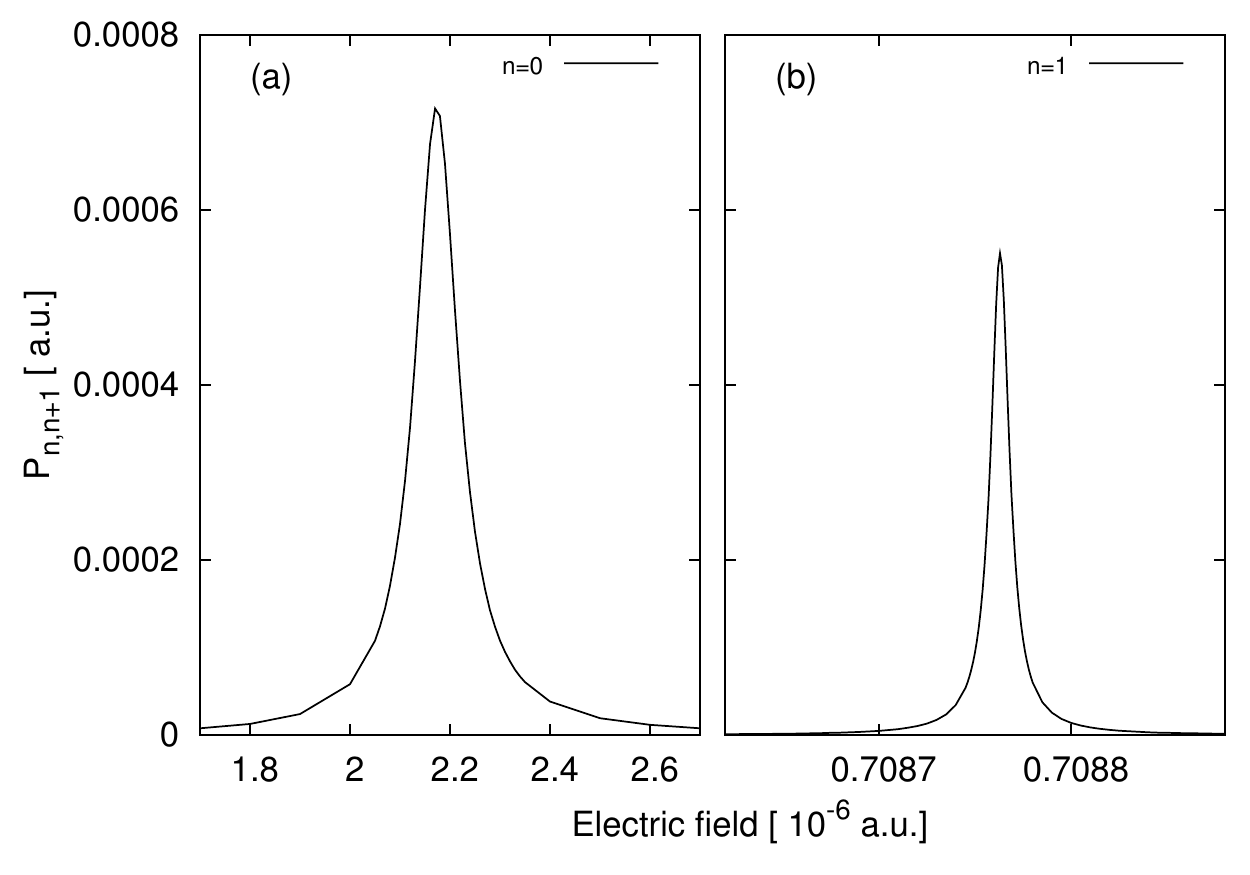} 
\caption{
The transition probability between the states $n$ and $n+1$ at the bottom of the
 conduction band in the system based on the Fibonacci sequence;
(a) $n=0$, (b) $n=1$.
}
\label{fig:13}
\end{figure}

\begin{figure}
\includegraphics[width=\columnwidth]{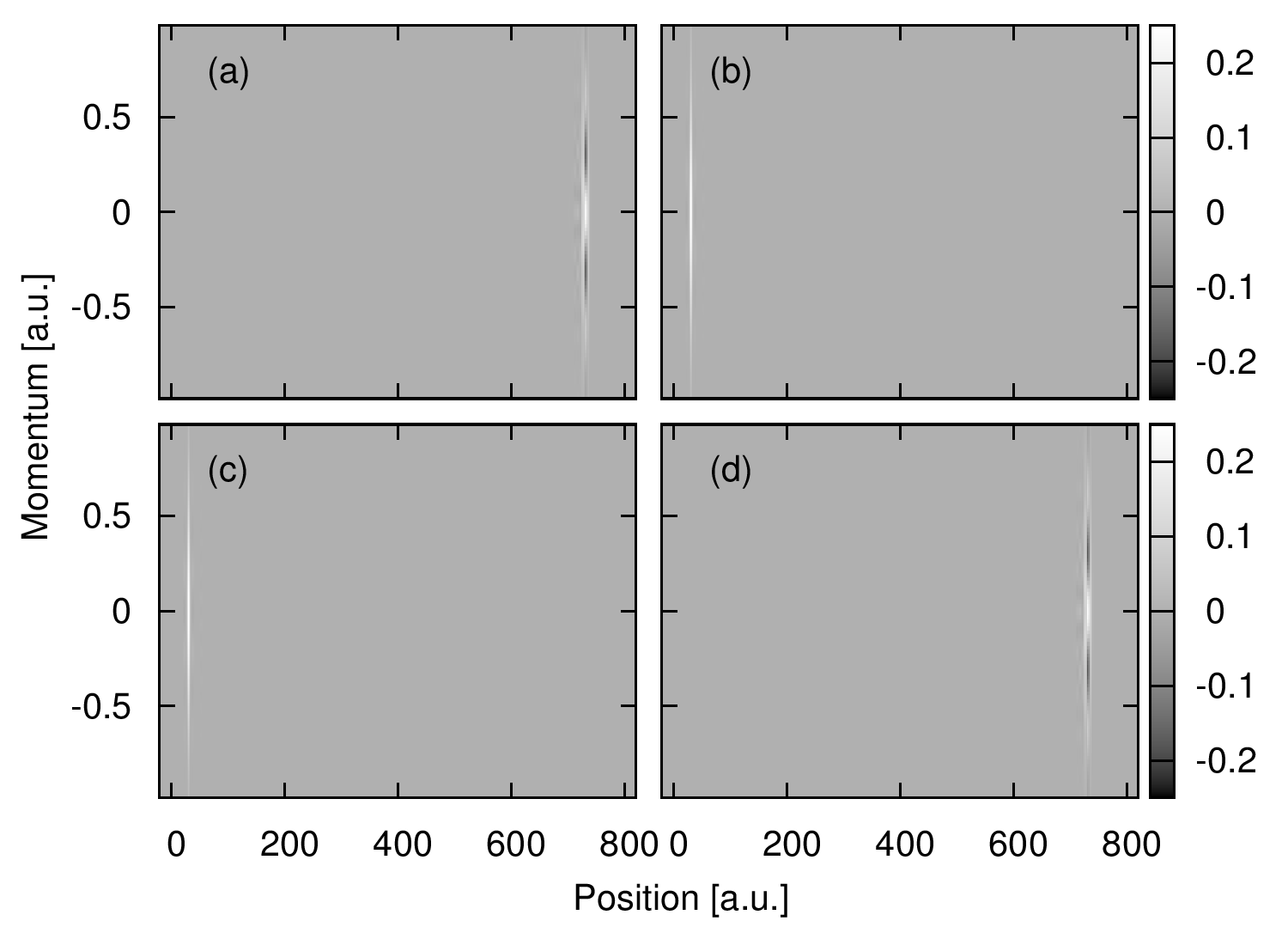} 
\caption{
The Wigner function for the system based on the Thue-Morse sequence.
(a), (b) -- $E_{F}$, (c), (d) -- $E_{F+1}$.
Electric field: (a) and (c)
0.98 $\times$ 10$^{-6}$ a.u.; (b) and (d) 0.99 $\times$ 10$^{-6}$ a.u.
}
\label{fig:06}
\end{figure}

Quantitative analysis of the transition between electronic states
$|\psi_n\rangle$ and $|\psi_m\rangle$ may be based on the product of their
Wigner distribution functions $f_n(x,k)$ and $f_m(x,k)$ integrated over the
phase space \cite{Dowling1991, Agarwal1995, Dragoman2005}, namely
\begin{equation}
P_{nm}=2\pi\int{\mathrm{d}x \, \mathrm{d}k}\;f_n(x,k)\,f_m(x,k).
\label{trans}
\end{equation}
The results of calculations of the overlapping integral given by
Eq.(\ref{trans}) for the three lowest states in the Fibonacci chain are shown in
Fig. \ref{fig:13}.
The maximum values of the overlapping integrals in both cases correspond to the
electric field values at which the anticrossings between the appropriate states
are observed.
After a further increase in the electric field the quantum interference term
disappears because the distance between states becomes larger and finally both
states are well separated in the phase space which means that the overlapping
integrals tend to zero.
A similar situation is observed for the electronic states in the Thue-Morse
 chain, for example in case of the Wigner functions shown in Fig. \ref{fig:06}.

\begin{figure}
\includegraphics[width=\columnwidth]{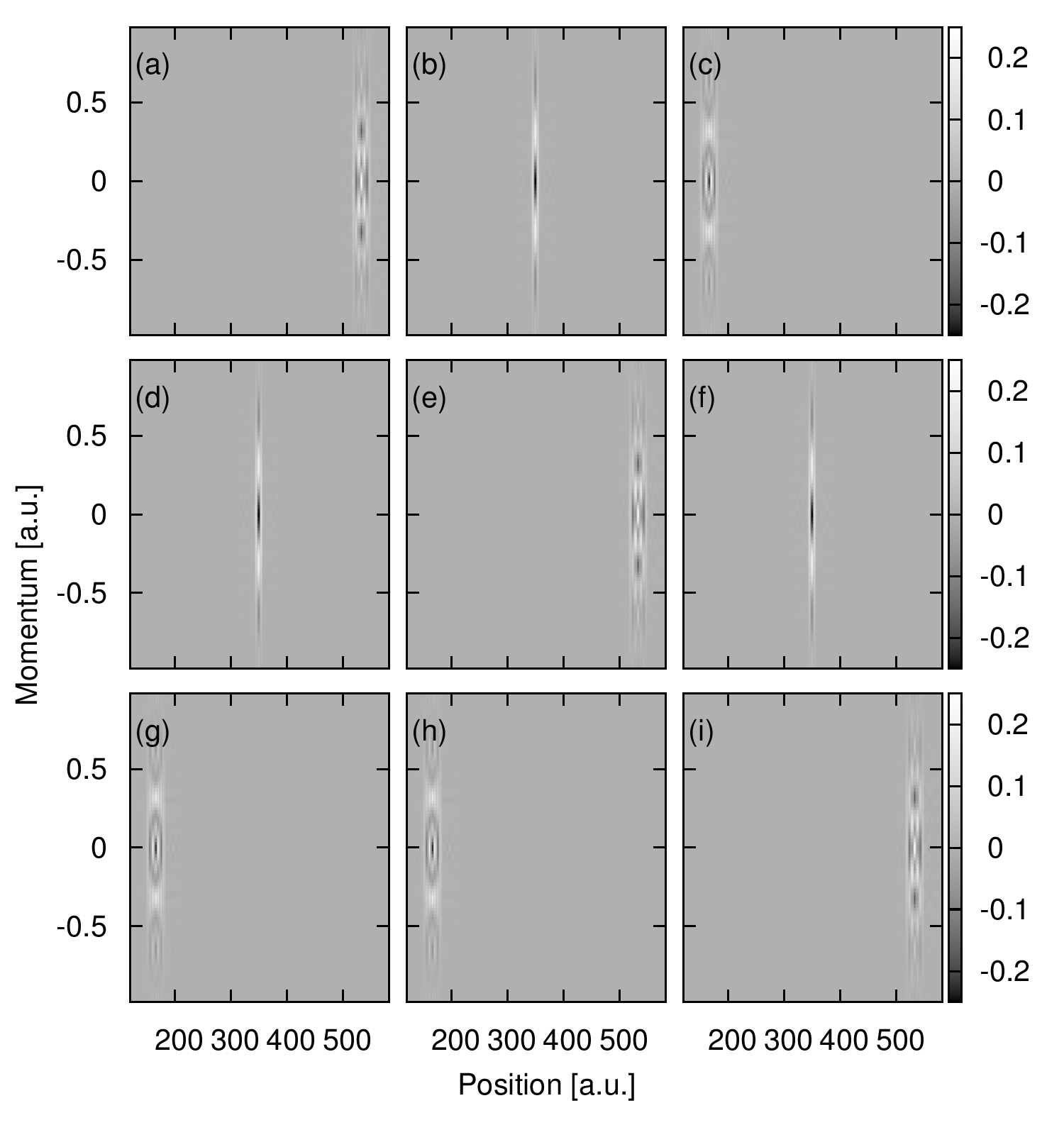} 
\caption{
The Wigner function for the system based on the Fibonacci sequence.
(a), (b), (c) -- $E_{F-1}$, (d), (e), (f) -- $E_{F}$, (g)-(i) -- $E_{F+1}$.
Electric field, from left to right, 1.8, 1.85, 1.9 $\times$ 10$^{-6}$ a.u.}
\label{fig:05}
\end{figure}

A more complex situation is presented in Fig. \ref{fig:05} where the evolution
of the Wigner distribution functions for three quantum states in the vicinity of
the middle of the conduction band is shown as a function of the electric field.
In this case the energy spectrum (see Fig. \ref{fig:02}) exhibits two
anticrossings located close to each other.
Each of these anticrossings mixes only two neighbouring electronic states,
therefore this process can be explained by the previous analysis.
As it is presented in Figs \ref{fig:04}, \ref{fig:06}, \ref{fig:05} quantum
states in the aperiodic chains under the fixed boundary conditions
localise in limited areas of the phase space.
The extents of these areas depend on the assumed distribution of the potential
wells (based on Fibonacci or Thue-Morse sequence) and the number of quantum
state.

One of the possibilities of measuring the degree of localisation is to use the
 IPR parameter calculated in phase space, defined
 by~\cite{Ingold2003,Woloszyn2004}
\begin{equation}
\mathrm{IPR}_n = \frac{1}{2\pi} \int {{\mathrm{d}x \, \mathrm{d}k}} \;
\big[g_n(x,k;1/2)\big]^2,
\label{ipr:def}
\end{equation}
 where $g_n(x,k;1/2)$ is the Husimi function.
The Husimi function is an example of the non-negative non-classical distribution
 functions which can be obtained by the convolution of the Wigner distribution
 function and a window function~\cite{Qian2004}
\begin{equation}
g_n(x,k;\Delta_{xk}) =
 \int {\mathrm{d}x^{\prime} \mathrm{d}k^{\prime}} \;
 W(x-x^{\prime}, k-k^{\prime}; \Delta_{xk})
 f_n(x^{\prime},k^{\prime}),
\label{splot:def}
\end{equation}
 where $W(x-x^{\prime}, k-k^{\prime}; \Delta_{xk})$ is a window function
 with resolution $\Delta_{xk}$.
In particular, the Husimi function is obtained if we choose the window function
 as a Gaussian function with the resolution corresponding to the minimum
 resulting from the uncertainty principle ($\Delta_{xk}=1/2$).

\begin{figure}
\includegraphics[width=\columnwidth]{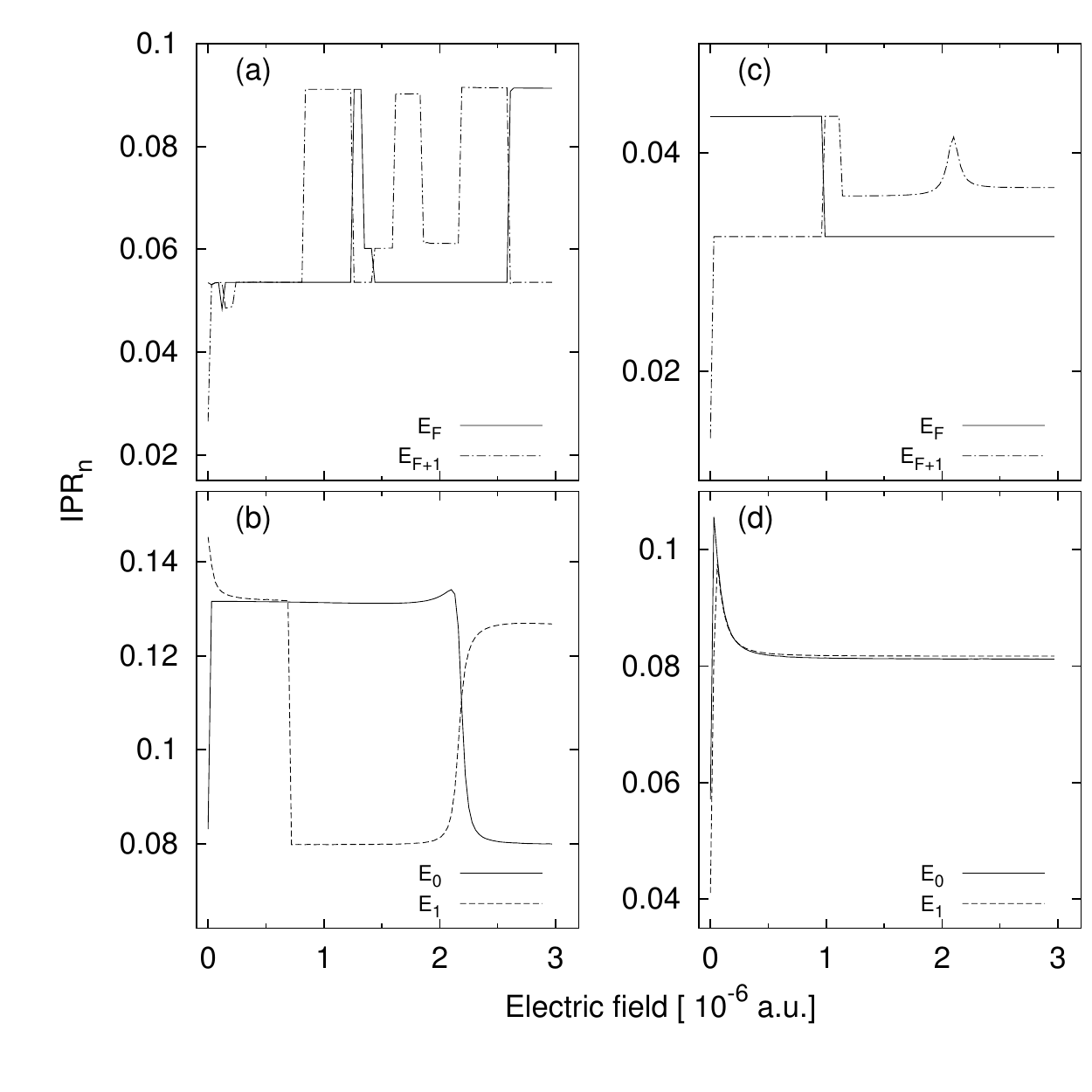} 
\caption{
Inverse Participation Ratio ( $\mathrm{IPR}_n$) for (a), (b) the Fibonacci sequence
 and (c), (d) Thue-Morse sequence.}
\label{fig:11}
\end{figure}

\begin{figure}
\includegraphics[width=\columnwidth]{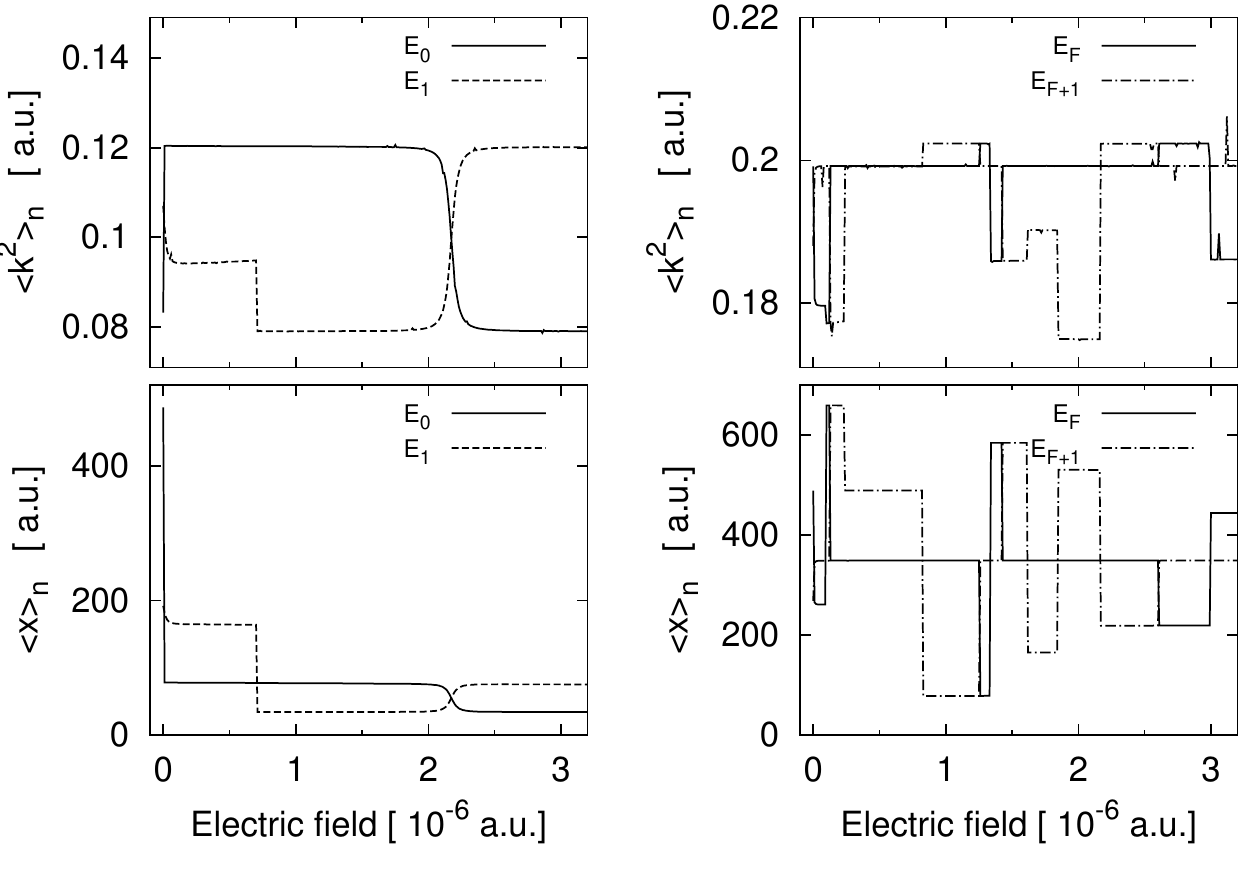} 
\caption{
The average position $\langle x \rangle_n$ and squared momentum
$\langle k^2 \rangle_n$ for the Fibonacci sequence.}
\label{fig:07}
\end{figure}

\begin{figure}
\includegraphics[width=\columnwidth]{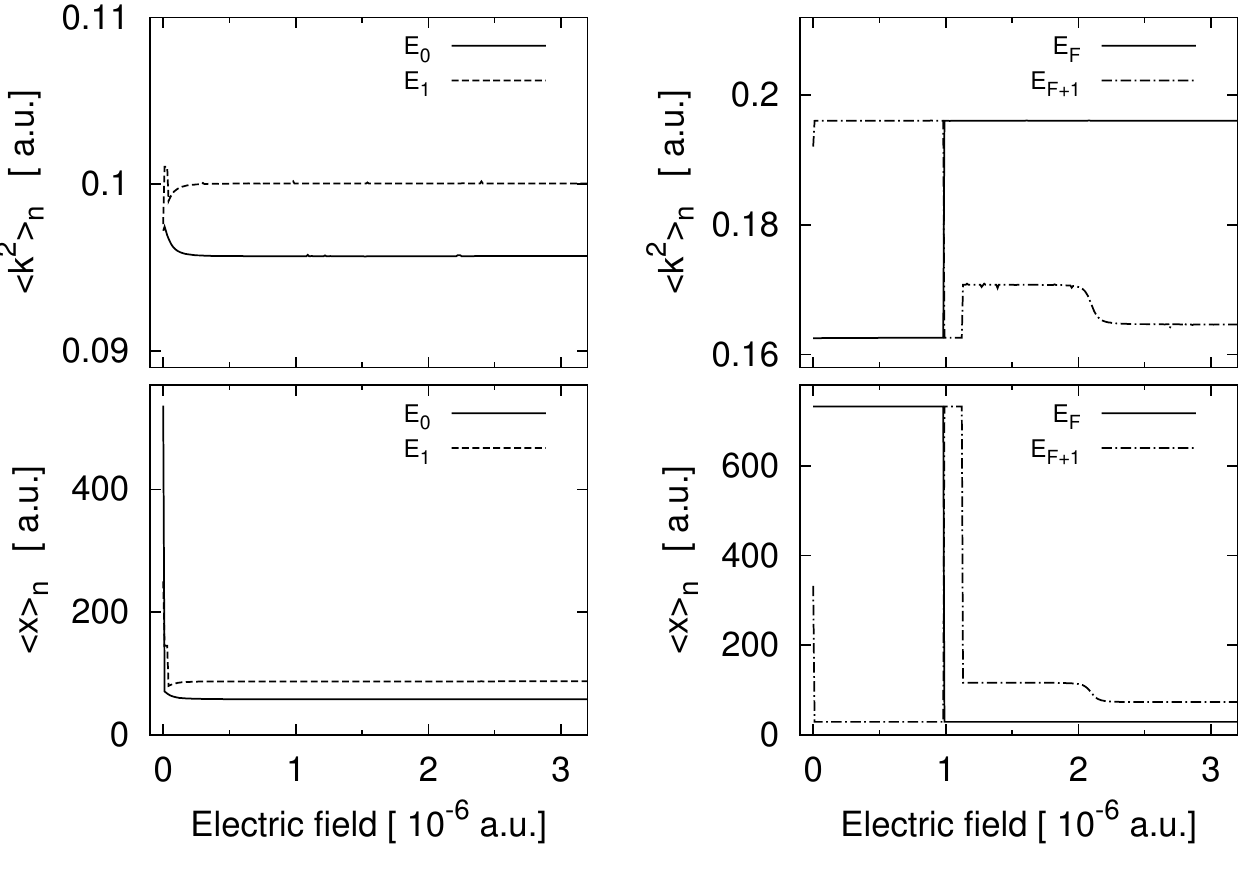} 
\caption{
The average position$\langle x \rangle_n$ and squared momentum
 $\langle k^2 \rangle_n$ for the Thue-Morse sequence.
}
\label{fig:08}
\end{figure}

IPR can be used to study the influence of the electric field on localisation,
 as shown in Fig.~\ref{fig:11}.
For the ground state in the Fibonacci chain (see Fig. \ref{fig:11}. b)  it was
found that after IPR reaches the value of 0.13, further increase in the
 electric field to $2 \times 10^{-6}$~a.u. does not change the localisation.
In a similar way, for the first excited state IPR stabilises initially at 0.08
for the fields larger than $0.7 \times 10^{-6}$~a.u.
Then, when the electric field is between 2 and $2.5\times 10^{-6}$~a.u.,  we
observe that the energy levels values plotted versus the electric field
 change their gradient (see Fig. \ref{fig:02}. b).
This behaviour is connected with the anticrossing taking place between the
mentioned values of the electric field.
In the phase space, the shape of the states changes, as well as the occupied
area.
Both states are shifted and the degree of localisation changes significantly.
As a consequence, the first excited state localises in the region of the phase
space previously occupied by the ground state, and vice versa.
The expectation values of the position and squared momentum presented on
Figs \ref{fig:07} and \ref{fig:08} were calculated for these states.
The results confirm that the electric field shifts the states only at the points
 of anticrossings.

In case of the Thue-Morse sequence the observed behaviour is slightly different.
When the electric field is applied, localisation of the two lowest states starts
 to increase very quickly and the states are shifted.
Then, the degree of localisation falls down and finally settles at a constant
level.
Also the expectation values of position and momentum do not undergo any
further changes when the electric fields are increased.
Similar characteristics of the ground state and the first excited state mean
that increasing the electric field separates energies of the states,
 as shown in Fig. \ref{fig:02}. d.

Figs \ref{fig:11}. a, c present changes in localisation of the states from the
middle of the band.
The changes are very sharp and are observed for the electric field values at
which the plot of energy of the state against the electric field alters its
gradient.
Modification of the phase space areas occupied by the states observed at the
anticrossings, together with the change in IPR values, are caused by the
 decreasing of the energy between the states.
For that reason we expect the quantum effects to have a greater impact on the
analysed states, notably the quantum interference between them.
Appearance of that kind of effects should be accompanied by an increase in the
negative part of the Wigner function which is  responsible for the quantum
interference.

\begin{figure}
\includegraphics[width=\columnwidth]{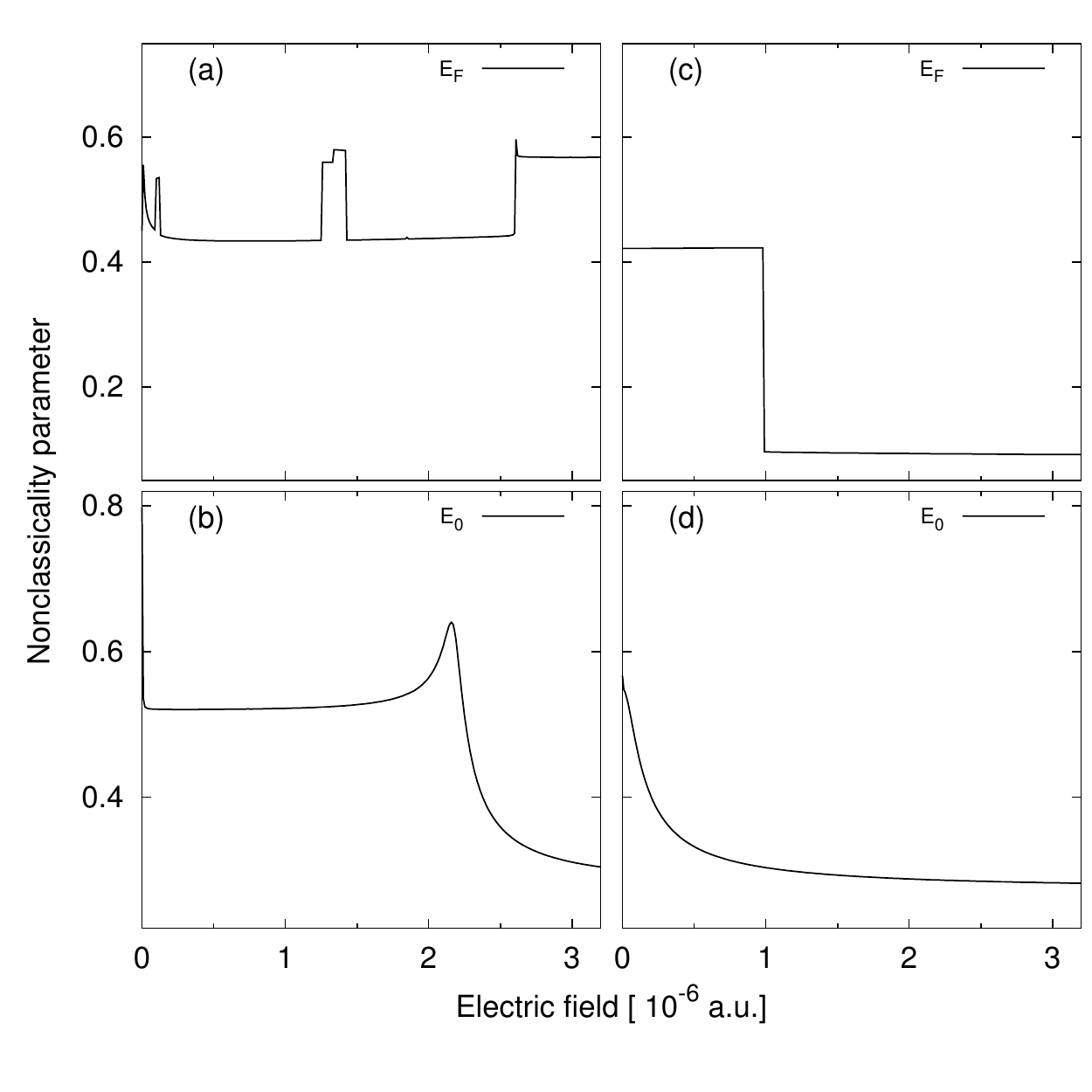} 
\caption{
The nonclassicality parameter $\nu_n$ calculated for the ground state ($n=0$) in
 case of (b) the Fibonacci sequence and (d) Thue-Morse sequence.
The nonclassicality parameter $\nu_n$ calculated for the states corresponding
with  the Fermi level in case of (a) the Fibonacci sequence and (c) Thue-Morse
sequence.}
\label{fig:03}
\end{figure}

To investigate this phenomenon we split up the Wigner distribution function into
two parts, namely
\begin{equation}
f_n(x,k)=f_n^{+}(x,k)+f_n^{-}(x,k),
\label{split}
\end{equation}
where $f_n^{+}(x,k)$ and $f_n^{-}(x,k)$ correspond to the positive part and the
negative part of the Wigner distribution function, respectively.\\
The measure of the non-classical nature of an electronic state $|\psi_n\rangle$
 is defined by the formula \cite{Benedict1999, Kenfack2004}
\begin{equation}
\nu_n = 1 -
\frac{\mathcal{I}_n^{+}-\mathcal{I}_n^{-}}{\mathcal{I}_n^{+}+\mathcal{I}_n^{-}},
\label{def:ncls}
\end{equation}
where $\mathcal{I}_n^{+}$ and $\mathcal{I}_n^{-}$ are the moduli of the
integrals of the positive part and the negative part of the Wigner distribution
function, respectively.
Influence of the electric field on the $\nu_n$ parameter is shown  in
Fig. \ref{fig:03}.\\
In the Fibonacci chain the non-classical character considerably decreases (to
about 0.52) immediately after turning the electric field on.
Then its value does not change until the anticrossing takes place in the energy
spectrum, where the $\nu_n$ parameter rises as a result of
 the increasing Wigner function negative part for both states.
For a further increase in electric field the non-classical nature decreases.\\
For the ground state in the Thue-Morse chain a monotonic decrease in the
parameter $\nu_n$ is observed.
It can be explained on the basis of large separation in the phase space between
the ground state and the excited states.\\
States placed in the middle of the band also exhibit jumps of the non-classical
behaviour at the electric fields at which the anticrossings are observed.
Here, however, the widths of anticrossings are very small, and the numerical
accuracy does not allow to illustrate the peaks of non-classicality parameter,
but only to show the values below and above anticrossings.

\begin{figure}
\includegraphics[width=\columnwidth]{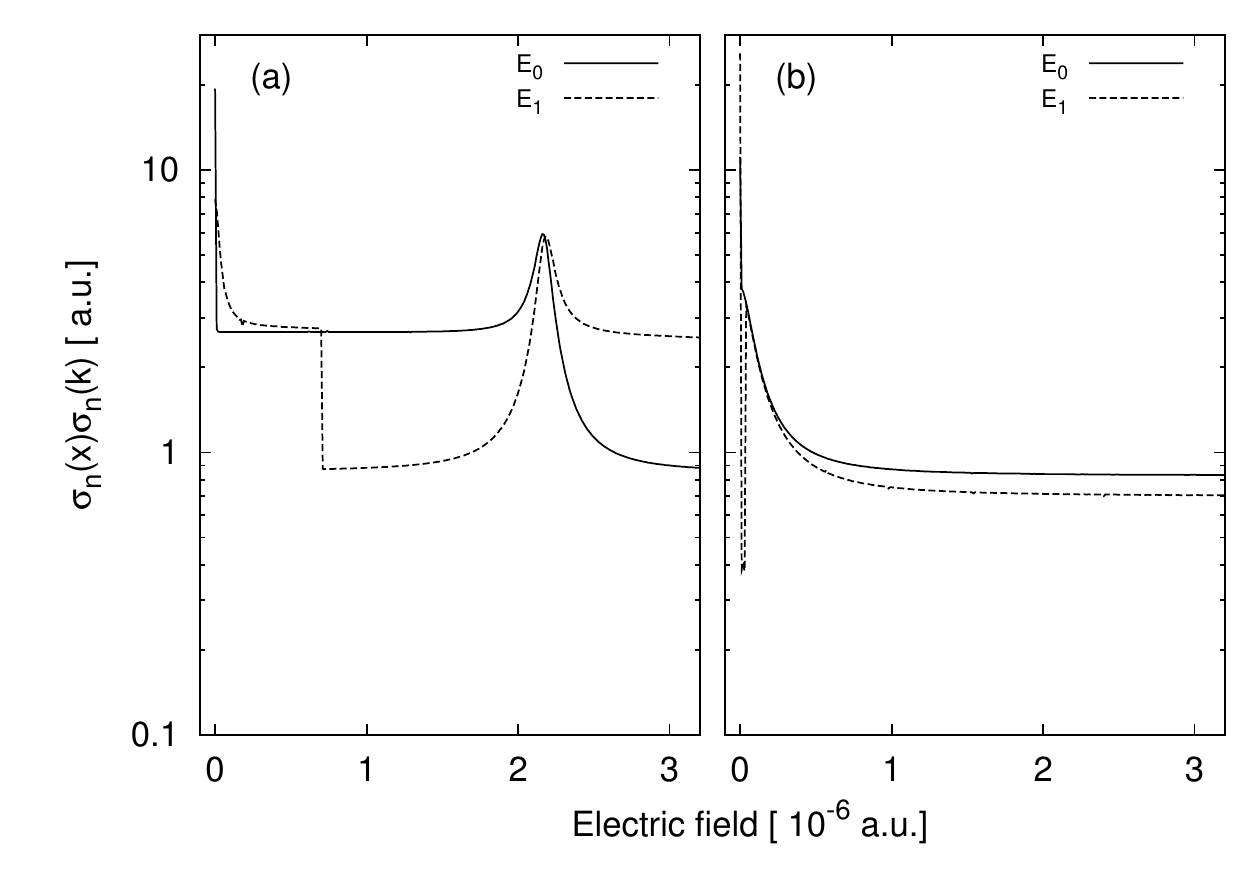} 
\caption{ The product of the position and momentum variances for the Fibonacci
sequence (a) and the Tue-Morse sequence (b).}
\label{fig:12}
\end{figure}

Additionally we calculate the standard deviations of position and momentum
variables that are defined by expressions: $\sigma_n^2(x)=\langle x^2
\rangle_n-\langle x\rangle_n^2$ and $\sigma_n^2(k)=\langle k^2\rangle_n-\langle
k\rangle_n^2$, respectively, applying Eq.~(\ref{eq:A_n}) to find the relevant
quantities from the Wigner function $f_n(x,k)$.
The acquaintance of these deviations allows one to calculate the uncertainty
product $\sigma_n(x) \sigma_n(k)$ as a function of the electric field.
Fig. \ref{fig:12} shows results for the uncertainty product for the two lowest
states in both aperiodic chains.
We may see that the electric field produces rapid changes in the uncertainty
product around the anticrossings for both states.
This behaviour of the uncertainty product can be explained by the highly
nonclassical features of the Wigner distribution function corresponding to the
quantum state in the region of anticrossing where the quantum interference
phenomena play important role.
On the other hand the uncertainty product allows one to distinguish between
pseudoclassical and nonclassical states.
It results from the fact that the uncertainty product is equal $1/2$ for the
coherent states which are recognised as the states closest to classical ones.
Hence the nonclassical states may be characterised by the uncertainty product
greater than $1/2$.
As it presented in Fig. \ref{fig:12}, the  uncertainty product is usually
greater than one which means that the discussed states are nonclassical.

\section{\label{sec:level4}Concluding remarks}

We have studied the influence of the homogeneous electric field on the energy
spectrum of the finite one-dimensional aperiodic systems generated by the
Fibonacci and Thue-Morse sequences.
We have shown that the homogeneous electric field modifies the energy spectrum
of aperiodic systems and leads to the formation of the subbands.
The appearance of subbands is a consequence of the attraction between energy
levels for some value of the electric field.
The anticrossings are observed for these values of the electric field.\\
We have applied the phase space method based on the Wigner distribution function
to analyse the region of anticrossings for the lowest electronic states in both
aperiodic systems.
The discussion of the electronic states in the middle of conduction band is also
included.
This formulation allows to investigate the nonclassical properties of the
electronic states and their influence on the transitions between them.
We have shown that the nonclassical properties of electronic states of the
aperiodic systems under the homogeneous electric field are most profound in the
regions of anticrossings.
The increase in the nonclassical parameter in these regions is a consequence of
the increasing role of the negative part of the Wigner distribution function and
it is correlated with the changes in the localisation properties of electronic
states and the dynamical variables.
Finally, we have used the uncertainty product of momentum and position variables
as a simple measure of nonclassicality of the electronic state.
We have found that this product as a function of the electric field is also
correlated with the nonclassical properties of electronic states in the
aperiodic systems.
In the paper we have paid attention to the aperiodic systems in the limit of low
 homogeneous electric field.
The limit of strong electric field and dynamical aspect of the problem will be
 included in forthcoming publications.

\begin{acknowledgments}
The authors thank G.J.~Morgan (University of Leeds) and A.Z.~Maksymowicz
 (AGH-UST) for valuable discussions.
This work was supported by the Polish Ministry of Science and Higher Education
 under the ''International scientist mobility support'' programme and
 by the AGH-UST under the project no. 11.11.220.01 ''Basic and applied
 research in nuclear and solid state physics''.
\end{acknowledgments}


\begin{thebibliography}{10}

\bibitem{Eigler1990}
D. M. Eigler and E. K. Schweizer,
Nature {\bf 344}, 524 (1990).

\bibitem{Meyer2000}
G. Meyer, and et al.,
Single Mol. {\bf 1}, 79 (2000).

\bibitem{Muller1992}
C.J. Muller, J.M. van Ruitenbeek, and L.J. de Jongh, Physica C {\bf 191}, 485
(1992).

\bibitem{Yanson1998}
A. I. Yanson, G. Rubio-Bollinger, H. E. van den Brom, N. Agrait, and J. M. van
Ruitenbeek,
Nature (London) {\bf 395}, 783 (1998).

\bibitem{Smit2001}
R. H. M. Smit, C. Untiedt, A. I. Yanson, and J. M. van Ruitenbeek,
Phys. Rev. Lett. {\bf 87}, 266102 (2001).

\bibitem{Ohnishi1998}
H. Ohnishi, V. Kondo, and K. Takayanagi,
Nature {\bf 395}, 780 (1998).

\bibitem{Bettini2006}
J. Bettini, F. Sato, P. Z. Coura, S. O. Dantas, D. S. Galvao, and D. Ugarte,
Nature Nanotechnology {\bf 1}, 182 (2006).

\bibitem{Crain2004}
J. N. Crain, J. L. McChesney, F. Zheng, M. C. Gallagher, P. C. Snijders, M.
Bissen, C. Gundelach, S. C. Erwin, and F. J. Himpsel,
Phys. Rev. B {\bf 69}, 125401  (2004).

\bibitem{Krawiec2006}
M. Krawiec, T. Kwapi\'nski, and M. Ja{\l}ochowski,
Phys. Rev. B {\bf 73}, 075415  (2006).

\bibitem{qts:lifszyc}
B. L. Altshuler, A. G. Aronov, D. E. Khmelnitskii, and A. I. Larkin
Coherent effects in disordered conductors,
MIR Publishers,
Moscow (1982)
(In: Quantum Theory of Solids, ed. I. M. Lifshits).

\bibitem{Lee1985}
P. A. Lee and T. V. Ramakrishnan,
Rev. Mod. Phys. {\bf 57}, 287 (1985).

\bibitem{kramer1993}
B. Kramer and A. MacKinnon,
Rep. Prog. Phys. {\bf 56}, 1469 (1993).

\bibitem{Wannier1960}
G. H.Wannier,
Phys. Rev. {\bf 117}, 432 (1960).

\bibitem{Avron1982}
J. E. Avron,
Ann. Phys. {\bf 143}, 33 (1982).

\bibitem{Nenciu1991}
G. Nenciu,
Rev. Mod. Phys. {\bf 63}, 91 (1991).

\bibitem{Grecchi1995}
V. Grecchi and A. Sacchetti,
Ann. Phys. {\bf 241}, 258 (1995).

\bibitem{Mendez1988}
E. E. Mendez, F. Agull\'o-Rueda, and J. M. Hong,
Phys. Rev. Lett. {\bf 60}, 2426  (1988).

\bibitem{Voisin1988}
P. Voisin, J. Bleuse, C. Bouche, S. Gaillard, C. Alibert, and A. Regreny,
Phys. Rev. Lett. {\bf 61}, 1639 (1988).

\bibitem{Ghulinyan2005}
M. Ghulinyan, C. J. Oton, Z. Gaburro, L. Pavesi, C. Toninelli, and D. S. Wiersma,
Phys. Rev. Lett. {\bf 94}, 127401 (2005).

\bibitem{Shockley1972}
W. Shockley,
Phys. Rev. Lett. {\bf 28}, 349 (1972).

\bibitem{Fukuyama1973}
H. Fukuyama,
Phys. Rev. B {8}, 5579 (1973).

\bibitem{Roy1982}
C. L. Roy and P. K. Mahapatra,
Phys. Rev. B {\bf 25}, 1046 (1982).

\bibitem{Soukoulis1983}
C. M. Soukoulis, J. V. Jose, E. N. Economou, and P. Sheng,
Phys. Rev. Lett. {\bf 50}, 764 (1983).

\bibitem{Prigodin1989}
V. N. Prigodin and B. L. Altshuler,
Phys. Lett. A {\bf 137}, 301 (1989).

\bibitem{Ryu1993}
C. S. Ryu, G. Y. Oh, and M. H. Lee, Phys. Rev. B 48, 132 (1993).

\bibitem{deBrito1995}
P. E. de Brito, C. A. A. da Silva, and H. N. Nazareno,
Phys. Rev. B {\bf 51}, 6096 (1995).

\bibitem{Farchioni1997}
R. Farchioni and G. Grosso,
Phys. Rev. B {\bf 56}, 1981 (1997).

\bibitem{Gluck1998}
M. Gl\"uck, A. R. Kolovsky, H. J. Korsh, and N. Moiseyev,
Eur. Phys. J. D. {\bf 4}, 239 (1998).

\bibitem{Rosam2001}
B. Rosam, D. Meinhold, F. L\"oser, V. G. Lyssenko, S. Glutsch, F. Bechstedt, F.
Ross, K. K\"ohler, and K. Leo,
Phys. Rev. Lett. {\bf 86}, 1307 (2001).

\bibitem{Gluck2002a}
M. Gl\"uck, A. R. Kolovsky, and H. J. Korsh,
Phys. Rep. {\bf 366}, 103 (2002).

\bibitem{Kast2002}
M. Kast, C. Pacher, G. Strasser, E. Gornik, and W. S. M. Werner,
Phys. Rev. Lett. {\bf 89}, 136803 (2002).

\bibitem{Zhang2006}
A. Zhang, L. C. Lew Yan Voon, and M. Willatzen,
Phys. Rev. B {\bf 73}, 045316 (2006).

\bibitem{Tatarskii83}
V. I. Tatarskii,
Sov. Phys. Usp. {\bf 26}, 311 (1983).

\bibitem{Hillery84}
M. Hillery, R. F. O'Connell, M. O. Scully, and E. P. Wigner.
Phys. Rep. {\bf 106}, 121 (1984).

\bibitem{Lee95}
H. -W. Lee,
Phys. Rep. {\bf 259}, 147 (1995).

\bibitem{Schleich2001}
W. P. Schleich, Quantum Optics in Phase Space, Wiley VCH, Weinheim (2001).

\bibitem{Luck1989}
J.M. Luck, Phys. Rev. B {\bf 39}, 5834 (1989).

\bibitem{Bovier1993}
A. Bovier, J.-M. Ghez,
Commun. Math. Phys. {\bf 158}, 45 (1993).

\bibitem{Hu2000}
B. Hu, B. Li, and P. Tong, Phys. Rev. B {\bf 61}, 9414 (2000).

\bibitem{Morgan1985}
G. J. Morgan, M. Howson, and K. Saub,
J. Phys. F: Metal. Phys. {\bf 15}, 2157 (1985).

\bibitem{Frensley1990}
W. R. Frensley,
Rev. Mod. Phys. {\bf 62}, 745 (1990).

\bibitem{Biegel1996}
B. A. Biegel and J. D. Plummer,
Phys. Rev. B {\bf 54}, 8070 (1996).

\bibitem{Jacoboni2004}
C. Jacoboni and P. Bordone,
Rep. Prog. Phys. {\bf 67}, 1033 (2004).

\bibitem{Wojcik2008}
P. W\'ojcik, B. J. Spisak, M. Wo{\l}oszyn, and J. Adamowski,
Acta Phys. Pol. A {\bf 114}, 933 (2008).

\bibitem{Spisak2007}
B. J. Spisak and M. Wo{\l}oszyn,
Acta Phys. Pol. B {\bf 38}, 1951 (2007).

\bibitem{Ingold2003}
G.-L. Ingold, A. Wobst, Ch. Aulbach, P. H\"angi, Lect. Notes Phys. 630, 85
(2003).

\bibitem{Woloszyn2004}
M. Wo{\l}oszyn, B. J. Spisak, Materials Science, Poland 22, 545 (2004).

\bibitem{Wilkinson1996}
S. R. Wilkinson, C. F. Bharucha, K. W. Madison, Q. Niu, and M. G. Raizen,
Phys. Rev. Lett. {\bf 76}, 4512 (1996).

\bibitem{Madison1999}
K. W. Madison, M. C. Fischer, and M. G. Raizen,
Phys. Rev. A {\bf 60}, R1767 (1999).

\bibitem{Billy2008}
J. Billy, V. Josse, Z. Zuo, A. Bernard, B. Hambrecht, P. Lugan, D. Clement, L.
Sanchez-Palencia, Ph. Bouyer, and A. Aspect,
Nature (London) {\bf 453}, 891 (2008).

\bibitem{Perez-Alvarez2001}
R. Perez-Alvarez, F. Garcia-Moliner, and V. R. Velasco,
J. Phys.: Condens. Matter {\bf 13}, 3689 (2001).

\bibitem{Gluck2002b}
M. Gl\"uck, A. R. Kolovsky, H. J. Korsch, and F. Zimmer,
Phys. Rev. B {\bf 65}, 115302 (2002).

\bibitem{Macia2006}
E. Macia,
Rep. Prog. Phys. {\bf 69}, 397 (2006).

\bibitem{Wannier1962}
G. H. Wannier,
Rev. Mod. Phys. {\bf 34}, 645 (1962).

\bibitem{Davydov1965}
A. S. Davydov, Quantum Mechanics 1st Edition, Pergamon Press, Oxford (1965).

\bibitem{Zurek1991}
W. Zurek,
Phys. Today {\bf 44}, 36 (1991) and arXive:quant-ph/0306072.

\bibitem{Benedict1999}
M. G. Benedict and A. Czirj\'ak,
Phys. Rev. A {\bf 60}, 4034 (1999).

\bibitem{Kenfack2004}
A. Kenfack and K. \.Zyczkowski,
J. Opt. B: Quantum Semiclass. Opt. {\bf 6}, 396 (2004).

\bibitem{Dowling1991}
J. P. Dowling, W. P. Schleich, and J. A. Wheeler,
Ann. Physik (Leipzig) {\bf 48}, 423 (1991).

\bibitem{Agarwal1995}
G. S. Agarwal,
Found. Phys. {\bf 25}, 219 (1995).

\bibitem{Dragoman2005}
D. Dragoman,
Phys. Scr. {\bf 72}, 290 (2005).

\bibitem{Qian2004}
Qian Shu Li, Gong Min Wei, and Li Qiang Lu,
Phys. Rev. A {\bf 70}, 022105 (2004).

\end{thebibliography}
\end{document}